\documentclass[a4paper,pre,amsmath,twocolumn,amssymb,unsortedaddress,showpacs]{revtex4-1}
\usepackage{graphicx}  
\usepackage{dcolumn}
\usepackage{bm}
\newcommand{\be}{\begin{equation}}
\newcommand{\ee}{\end{equation}}

\newcommand{\sech}{\mathrm{sech}}

\begin{document}

\title{Travelling  solitons in the externally driven 
 nonlinear Schr\"odinger equation}

 \author{I. V.  Barashenkov}
\email{Email: igor.barashenkov@gmail.com}
\affiliation{Department of  Mathematics,
University of Cape
Town, Rondebosch 7701;}
 \affiliation{National Institute for Theoretical Physics, Stellenbosch, South Africa}
 
\author{E. V.  Zemlyanaya}
\email{Email: elena@jinr.ru}
\affiliation{Joint Institute for Nuclear Research, Dubna, 141980 Russia}

\date{\today}

\begin{abstract}
We consider the undamped nonlinear Schr\"odinger equation 
driven by a periodic external force. 
Classes of travelling solitons and multisoliton complexes are obtained by the 
numerical continuation in the parameter space.
Two previously known  stationary solitons and two newly found localised solutions
are used as the starting points for the continuation.

We show that there are two families of stable solitons: one family
is stable for sufficiently
low velocities while solitons from the second family stabilise
when travelling faster than a certain critical speed.
The stable solitons of the former family 
can also form stably travelling bound states.
\end{abstract}


\maketitle

\section{Introduction}

The  damped nonlinear Schr\"odinger equation driven by a time-periodic  external  force,
\begin{subequations}
\be
iu_t+ u_{xx} + 2 |u|^2u + \delta u = a e^{i \Omega t} - i \beta u,
\label{u12}
\ee
and its parametrically driven counterpart 
 model  two fundamental energy supply mechanisms
in  a nearly-conservative spatially distributed system. 
While the  unperturbed  Schr\"odinger is an archetypal equation 
for the slowly varying envelope of a group of dispersive  waves,
the damped-driven  equations arise whenever the 
resonant forcing of small amplitude is used to compensate 
weak  dissipative losses.

The simplest (and perhaps the most visually appealing) realisation of Eq.\eqref{u12}
is that of the amplitude equation for 
a strongly coupled  pendulum array
with the horizontal
sinusoidal driving \cite{FK}, taken in its continuum limit.
 Here $a$ and $\Omega$ are the driving strength and driving frequency,
 respectively; $\delta$ is the detuning of the driving frequency from the 
 continuum of linear waves in the array, and $\beta$ is the damping coefficient.
 
The array of torsionally coupled pendula can serve as  a  prototype model for the whole
 variety of systems in condensed matter physics. Accordingly, 
Eq.\eqref{u12} was employed to study systems as diverse as
the ac-driven long
Josephson junctions \cite{Jj} and
charge-density-wave
conductors with external electric field \cite{CDW}; 
double-layer quantum Hall (pseudo)ferromagnets \cite{Hall}
and easy-axis ferromagnets in a
rotating magnetic field  \cite{magnetism}. 
 Eq.\eqref{u12}  arises 
 in the theory of rf-driven waves in
plasma \cite{plasma,NB_plasma} and shear
flows in nematic liquid crystals \cite{LC};
the same equation governs the amplitude of the slowly varying $\pi$-mode in the 
forced Fermi-Pasta-Ulam lattice \cite{FPU}.

A closely related equation is the one  with the {\it spatially\/} periodic forcing,
\begin{equation}
iu_t+ u_{xx} + 2 |u|^2u + \delta u = a e^{iKx} - i \beta u,
\label{u1}
\end{equation}
and, more generally, 
the one driven by the harmonic wave \cite{Cohen,trivial,driven_by_plane_wave}:
\be
iu_t+ u_{xx} + 2 |u|^2u + \delta u = a e^{i (Kx +\Omega t)} - i \beta u.
\label{u120}
\ee
\end{subequations}
A  discrete version of Eq.\eqref{u1} 
describes  an array of coupled-waveguide resonators excited by a driving field 
\cite{Egorov_Flach}
 whereas Eq.\eqref{u120}  models pulse propagation in 
an asymmetric twin-core optical fiber \cite{Cohen}. 

Equation \eqref{u120} includes \eqref{u12} and \eqref{u1}
as particular cases.
The transformation 
\[
u(x,t)=\Psi(X, t) e^{i(Kx+\Omega t) }, \quad  X=x-2Kt
\]
takes  \eqref{u120}  to 
\be
i \Psi_t + \Psi_{X X} + 2 |\Psi|^2 \Psi  - \kappa^2 \Psi= a- i \beta \Psi,
\label{auto}
\ee
with $\kappa^2= K^2+\Omega -\delta$.
The equation in this form has a history of applications of its own ---  in particular, in the physics of
optical cavities.  Originally, it was
introduced as the Lugiato -  Levefer model \cite{Lugiato_Lefever} 
of the diffractive cavity driven by a plane-wave stationary beam. Later it was employed to
describe a synchronously pumped ring laser with a nonlinear dispersive fiber \cite{fiber,Wabnitz}.
More recently the same equation was shown to govern the envelopes of short baroclinic Rossby waves
in the two-layer model of the atmosphere, or the ocean \cite{Rossby}.

Equation \eqref{auto} has undergone an extensive mathematical analysis. Topics covered included 
existence \cite{KN,BS1,BZ_PhysicaD},
 stability  \cite{BZB,BS1} and bifurcation \cite{NB_plasma,bifurcations}  of nonpropagating solitons
and their bound states \cite{Malomed,Wabnitz,BSA,Kollmann}; 
statistical mechanics of soliton creation and annihilation \cite{stats};
 soliton
autoresonance phenomena \cite{autoresonance,driven_by_plane_wave};
regular  \cite{Terrones} and 
chaotic \cite{chaos}  attractors on finite spatial intervals.
Here and below we use the word ``soliton" simply as a synonym 
for ``localised travelling wave".

The recent paper \cite{MQB} studied  solitons of the undamped ($\beta=0$)
 equation \eqref{auto} travelling
with constant or oscillating velocities.
Summarising results of their direct numerical simulations
of Eq.\eqref{auto}, the authors 
 formulated an empirical stability criterion of the soliton
 against small and large perturbations. 
So far, this criterion has not been given any mathematical proof or physical justification.
Despite being tested on
a variety of initial conditions, it still has the status of conjecture.

In order  to verify the validity of the empirical stability criterion
at least for infinitesimal perturbations, one needs to have
the travelling soliton existence and linearised stability domains accurately demarcated.
The classification of bifurcations
 occurring when stability is lost would also be a useful step towards 
the justification of  the criterion.
 This is what we shall concern ourselves with in this paper. 
 
Here, we study travelling solitons of Eq.\eqref{auto} by path-following them 
in the parameter space. One advantage of this approach over simulations is that it furnishes
{\it all\/} soliton solutions moving with a given velocity --- all stable and all unstable.
This, in turn, allows one
to
understand
the actual mechanisms
and details of the soliton transformations. 

The outline of this paper is as follows. 
In the next section, we give a brief classification of space- and time-independent
 solutions of Eq.\eqref{auto} which may serve as the backgrounds for the solitons.
 In particular, we show that there is only one stable background and 
 determine the value of the limit speed of the soliton propagating
 over it.
 In section \ref{Ins_lin} we describe insights one can draw from the analysis of the eigenvalues
 of the symplectic linearised operator and its hermitian counterpart. These pertain to
 the stability and bifurcation of the solitons.

 In section \ref{Nonpropagating} we present four {\it nonpropagating\/} directly driven solitons. 
 Two of these have already been available in literature
 while the other two have not been known before. 
 In sections \ref{Numerical_Simple} and \ref{Numerical_Twist},  we report on the continuation of
 these stationary solitons  to nonzero velocities.
 Our results on the existence and stability of the travelling
 solitons and their complexes, are summarised in 
 section \ref{Conclusions}. In particular, Fig.\ref{chart} gives a chart of ``stable" velocities 
 for each value of the driving strength.

\section{Flat solutions}
\label{Flat}

Assuming that $\kappa^2>0$  and defining 
$t'= \kappa^2 t$, 
$x'= \kappa  X$,  and
$\Psi= \kappa \psi$,
equation \eqref{auto} becomes
\[
i \psi_{t'} + \psi_{x' x'} + 2 |\psi|^2 \psi - \psi= -h -i \gamma \psi,
\]
where 
$h = -a/\kappa^3$,   $\gamma = \beta/\kappa^2$.
(In what follows, we omit primes above $x$ and $t$ for notational convenience.)

In this paper we study the above equation  with zero damping: $\gamma=0$.
Without loss of generality we can assume that $h>0$.
Since we shall be concerned with solitons travelling at nonzero 
velocities, it is convenient to transform the equation to a co-moving frame:
 \begin{equation}
i \psi_{t} -iV \psi_\xi+ \psi_{\xi \xi} + 2 |\psi|^2 \psi - \psi= -h,
 \label{our4}
 \end{equation}
where $\xi =x-Vt$.

Flat solutions are roots of the cubic equation
\begin{equation}
2|\psi|^2 \psi- \psi=-h;  \label{algebraic} 
\end{equation}
these have been classified in \cite{BS1}. 
 If $0<h<(2/27)^{1/2}$, there are 
3 roots, of which two ($\psi_1$ and $\psi_2$) are positive, and one ($\psi_3$)
is negative. Here
$\psi_1^2 < \frac16 <\psi_2^2< \frac12<\psi_3^2 <\frac23$.
If $h> (2/27)^{1/2}$, there is only one (negative) solution $\psi_3$, with
$\psi_3^2>\frac23$.

Let $\psi_0$ denote a root of equation \eqref{algebraic} --- one of the three roots $\psi_1$, $\psi_2$
and $\psi_3$. 
The value  $\psi_0$ does not  depend on $V$: the flat solution 
has the same form in any frame of reference. However
the spectrum of small perturbations of the flat solution does include a dependence
on $V$. 
Letting $\psi=\psi_0+ [u(\xi)+iv(\xi))]e^{\lambda t}$ in \eqref{our4}, 
linearising in $u$ and $v$, and, finally,  taking $u, v  \propto e^{ik \xi}$, we obtain
\begin{equation}
(\lambda - ikV)^2=-(k^2+a^2)(k^2+b^2),
\label{dispersion}
\end{equation}
where we have introduced 
\be
a= \sqrt{1-6\psi_0^2}, \quad b=  \sqrt{1-2 \psi_0^2}.
\label{ab}
\ee

To determine whether $\psi_0$ can serve as a background to
a stationary localised solution of \eqref{our4}, 
consider a time-independent perturbation --- that is, set $\lambda=0$:
\begin{equation}
k^2V^2=(k^2+a^2)(k^2+b^2).
\label{decay}
\end{equation}
The only  flat solution that  is 
{\it a priori\/} unsuitable as a background for 
 localised solutions is such $\psi_0$  whose  
 associated quadratic equation \eqref{decay}
has two nonnegative real roots, $(k^2)_1 \geq 0$ and $(k^2)_2 \geq 0$. 

It is not difficult to check that the negative solution $\psi_3$ 
has two nonnegative roots for any choice of $h$ and $V$.
This disqualifies $\psi_3$ as a possible soliton background.
We also conclude that travelling solitons may not exist for
$h$ greater than $(2/27)^{1/2}$.

Next, if $V \leq c$, where 
\be
c=a+b,
\label{c}
\end{equation}
the smaller positive solution $\psi_1$ will have either two
complex  or two negative roots $(k^2)_{1,2}$,
whereas  for velocities greater than $c$,  both roots are nonnegative.
Hence the $\psi_1$ solution can serve as a background only
for $V \leq c$. When $V< b-a$, the decay to the 
background is monotonic (both roots are negative),
while when $V>b-a$, the decay 
is by ondulation (the roots are complex). 
This flat solution admits a simple explicit expression: 
\[
\psi_1 = \sqrt{\frac23} \cos \left( \frac{\alpha}{3} - \frac{2 \pi}{3} \right),
\]
where
\[
\alpha = \arccos \left(- \sqrt{\frac{27}{2}} h \right),
\quad  \frac{\pi}{2} \leq  \alpha \leq \pi .
\]

Finally, the larger positive solution $\psi_2$ has two real roots 
of opposite signs (for all $V$ and $0 \leq h \leq (2/27)^{1/2}$).
This flat solution may also serve as a 
soliton background. 

Next, one can readily check that a flat solution $\psi_0$ is stable
if $\psi_0^2 < \frac16$. Therefore, even if there are
solitons asymptotic to the flat solution $\psi_2$ as $x \to \infty$ or
$x \to -\infty$, these will be of little physical interest as 
the background $\psi_2$ is always unstable.

In summary,  only the small positive flat solution (the one with $\psi_0^2< \frac16$) 
is  stable. It
may serve as a background for solitons  only if $V <c$; that is, the soliton propagation speed is limited by $c$.

The inequality $V \leq c$ limiting the soliton propagation speed, has a simple
physical interpretation. Indeed, one can easily check that $c$  gives the lower
bound for the phase velocity of radiation waves [in the original $(x,t)$ reference frame].
Therefore, a soliton travelling faster than $c$ would be exciting resonant radiation.
This is inconsistent with the asymptotic behaviour $\psi_x \to 0$ as $|x| \to \infty$;
neither could it be reconciled with the energy conservation.

\section{Insights from linearisation}
\label{Ins_lin}

Travelling wave solutions depend on $x$ and $t$ only in combination $\xi=x-Vt$.
For these, the partial differential equation \eqref{our4} reduces to
an ordinary differential equation
\begin{equation}
 -iV \psi_\xi+ \psi_{\xi \xi} + 2 |\psi|^2 \psi - \psi= -h.
\label{stationary}
\end{equation}
It is this  equation that we will be solving numerically in the following sections.

Let $\psi_s(\xi)$ be a localised solution of \eqref{stationary}. 
In order to represent results of continuation graphically, we will need to characterise 
the function $\psi_s(\xi)$ by a single value. A convenient choice for such a 
bifurcation measure is the momentum integral 
\begin{equation}
P= \frac{i}{2} \int (\psi^*_\xi \psi - \psi_\xi \psi^*) d \xi.
\label{P}
\end{equation}
One advantage of this choice is that 
the momentum is an integral of motion for equation \eqref{our4}; hence $P$ is a physically
meaningful characteristic of solutions. 
Another useful property of the momentum is that in some cases its extrema  mark the 
change of the soliton stability properties (see below). 

\subsection{The hermitian and symplectic operator}

Many aspects of the soliton's bifurcation diagram can be explained 
simply by the behaviour of the eigenvalues of the operator of linearisation about the 
travelling-wave solution in question. Therefore, 
before proceeding to the numerical continuation of travelling waves, we
introduce the linearised operator and discuss some of its properties.

Consider a 
perturbation of the solution of  Eq.\eqref{stationary}
 of the form $\psi= \psi_s +[u(\xi)+ iv(\xi)] e^{\lambda t}$, with small $u$ and $v$.
Substituting $\psi$ in Eq. \eqref{our4} and linearising in $u$ and $v$, we get a symplectic eigenvalue problem
\begin{equation}
\mathcal{H} {\vec y} = \lambda J {\vec y}.
\label{EV}
\end{equation}
Here ${\vec y}$ is a two-component vector-function
\[
{\vec y}(\xi)  = \left( \begin{array}{c} u  \\ v  \end{array} \right),
\]
and $\mathcal{H}$ is a  hermitian differential operator acting on such functions:
\[
\mathcal{H}= \left(
\begin{array}{cc}
-\partial_\xi^2 +1 -2(3 \mathcal{R}^2+ \mathcal{I}^2)   & -V \partial_\xi -4 \mathcal{R} \mathcal{I} \\
V \partial_\xi - 4 \mathcal{R} \mathcal{I} & 
-\partial_\xi^2 + 1 -2(3 \mathcal{I}^2 + \mathcal{R}^2) 
\end{array}
\right),
\]
with
$\mathcal{R}$ and $\mathcal{I}$ denoting the real and imaginary
part of the solution $\psi_s(\xi)$:  $\psi_s= \mathcal{R}+i \mathcal{I}$.
Finally, $J$ is a constant skew-symmetric matrix
\[
J = \left( 
\begin{array} {cc} 0 & -1 \\ 1 & 0 \end{array} \right).
\]

Assume that  $\psi_s(\xi)$ is a localised solution decaying to $\psi_0$ as $x \to \pm \infty$,
where $\psi_0^2< \frac16$.
The continuous spectrum of the hermitian
operator $\mathcal{H}$ occupies the positive real axis 
with a gap separating it from the origin:
$E \geq E_0 >0$. Discrete eigenvalues $E_n$ satisfy $E_n < E_0$. 
On the other hand, the continuous spectrum of the {\it symplectic\/} eigenvalues
(that is, the continuous spectrum of the operator $J^{-1} \mathcal{H}$)
occupies the imaginary axis of $\lambda$ outside the gap $(-i \omega_0, i \omega_0)$.
The gap width here is given by 
\be
\omega_0= \sqrt{(k_0^2+ a^2)(k_0^2+b^2)}-Vk_0 >0,
\label{omega0}
\ee
where $k_0$ is the positive root of the bicubic equation
\[
V^2(k^2+a^2)(k^2+b^2)= k^2(2k^2+ a^2+b^2)^2.
\]
Discrete eigenvalues of the operator $J^{-1} \mathcal{H}$ may include pairs of opposite
real values $ \lambda= \pm \rho$; 
 pure imaginary pairs
$\lambda= \pm i \omega$, with $0 \leq \omega \leq \omega_0$; and, finally,  complex
quadruplets $\lambda= \pm \rho \pm i \omega$. 

We routinely evaluate the spectrum of symplectic eigenvalues as we 
continue localised solutions in $V$. 
 If there is at least one eigenvalue $\lambda$ 
with $\textrm{Re} \lambda>0$, the solution $\psi_s$ is considered linearly unstable. 
Otherwise (that is, if all eigenvalues have $\textrm{Re} \lambda \leq 0$), the 
solution is deemed linearly stable. 

\subsection{Zero eigenvalues}

While the eigenvalues  of the operator $J^{-1} \mathcal{H}$
(that is,  the eigenvalues  of the symplectic eigenvalue problem \eqref{EV})
determine stability or instability of the solution $\psi_s$, the eigenvalues of the operator $\mathcal{H}$
are significant for the continuability of this solution.  Of particular importance are its zero eigenvalues.

At  a generic point $V$, the operator $\mathcal{H}$ has only one zero eigenvalue,
with the translational eigenvector ${\vec \Psi}_\xi \equiv ( \mathcal{R}_\xi, \mathcal{I}_\xi)$.
This is due to the fact that the stationary equation \eqref{stationary}
has only one continuous symmetry.
For a given $V$, the solution $\psi_s(\xi)$ is a member of a {\it one}-parameter family
of solutions $\psi_s(\xi-\theta)$, where $\theta$ is an arbitrary translation. 

On the other hand,  the nonhermitian operator 
$J^{-1} \mathcal{H}$ has two zero eigenvalues
at  a generic point. The reason is that the
equation \eqref{our4} as well as its linearisation,  are hamiltonian systems.
Real and imaginary eigenvalues of  operators which generate
hamiltonian flows always come in  pairs: If $\mu$ is an eigenvalue, so is $-\mu$
\cite{Arnold}.  
The two zero eigenvalues of the operator $J^{-1} \mathcal{H}$  reflect the fact that 
the function $\psi_s(\xi)$, considered as a solution of the partial differential
equation \eqref{our4}, is a member of a two-parameter family.  One parameter
is the translation; the other one is the velocity $V$.

For generic $V$, the  repeated zero eigenvalue of $J^{-1} \mathcal{H}$ is defective:
 there is only one eigenvector ${\vec \Psi}_\xi$ associated with it. 
 There is also a generalised eigenvector ${\vec \Psi}_V$, 
 where
 \[
{\vec \Psi}_V \equiv  \left(\frac{\partial \mathcal{R}}{ \partial V}, \frac{\partial \mathcal{I}}{\partial V} \right).
\]
This vector-function is {\it not} an eigenvector of $J^{-1} \mathcal{H}$;
instead, differentiating \eqref{stationary} in $V$ one checks that ${\vec z}={\vec \Psi}_V$  satisfies the nonhomogeneous equation
\be
\mathcal{H} {\vec z} = -J {\vec \Psi}_\xi.
\label{HVJ}
\ee
[That is,  ${\vec \Psi}_V$
is an eigenvector of the {\it square\/} of the symplectic operator:
$(J^{-1} \mathcal{H} )^2{\vec \Psi}_V=0 $.]

As we continue in $V$, a pair of opposite pure-imaginary symplectic eigenvalues may
collide at the origin on the $\lambda$-plane and cross to the positive
and negative real axis, respectively. The algebraic multiplicity of the 
 eigenvalue $\lambda=0$ increases from 2 to 4 at the point $V=V_c$;
however if the hermitian operator $\mathcal{H}$ does not acquire the second eigenvalue $E=0$
at this point, the geometric multiplicity remains equal to 1.
The change of stability of the soliton solution
does not affect its continuability, i.e. the soliton exists on either side  of $V=V_c$. 
In this case we have $dP/dV=0$ at the point where the stability changes \cite{Baer}.

The continuation may be obstructed only when another (the second) eigenvalue of the operator $\mathcal{H}$ 
crosses through zero at $V=V_c$:
$\mathcal{H} {\vec \Phi}=0$. 
If the corresponding eigenvector ${\vec \Phi}$ is not
orthogonal to the vector-function $J {\vec \Psi}_\xi$ in the right-hand side of  equation \eqref{HVJ},
its solution ${\vec z}= {\vec \Psi}_V$ will not be bounded. 
This implies a saddle-node bifurcation; the soliton solution $\psi_s$ cannot be continued beyond $V=V_c$. 
Note that although  ${\vec \Phi}$ is 
an eigenvector of the symplectic operator $J^{-1} \mathcal{H}$, the algebraic multiplicity
of the symplectic eigenvalue remains equal to 2
in this case.

Assume now that the eigenvector ${\vec \Phi}$ {\it is\/}
orthogonal to  $J {\vec \Psi}_\xi$.
This may  happen if  the soliton solution $\psi_s$ of 
equation \eqref{stationary} with  $V=V_c$
is a member of a {\it two}-parameter family of
solutions  $\psi_s=\psi_s(\xi-\theta; \chi)$,
with $\chi$ equal to some $\chi_0$.
Here we assume that  {\it each\/} member of the family
$\psi_s(\xi-\theta; \chi)$ is a solution of Eq.\eqref{stationary} --- with  the same $V=V_c$.
Then ${\vec \Phi}$  is given by
${\vec \Psi}_\chi \equiv \left. \partial {\vec \Psi}/ \partial \chi \right|_{\chi=\chi_0}$. 
If $\chi_0$ is a root of the equation 
\begin{subequations} \label{F}
\be
F(\chi)=0, 
\label{F1}
\ee
where
\be
 F(\chi) \equiv
\int ({\vec \Psi}_\chi,   J {\vec \Psi}_\xi) \, d \xi,
\label{F2}
\ee
\end{subequations}
the vectors ${\vec \Psi}_\chi$ and $J {\vec \Psi}_\xi$ will be orthogonal
which, in turn, will imply that a bounded solution ${\vec \Psi}_V$ of the equation  \eqref{HVJ} exists.
[In Eq.\eqref{F2} $(\phantom{a}, \phantom{b})$ stands for the
$\mathbb{R}^2$ scalar product: $({\vec a}, {\vec b}) \equiv a_1 b_1+ a_2 b_2$.]
In this case the value $V_0$ is {\it not\/} a turning point; the soliton solution $\psi_s$
exists on both sides of $V=V_0$.
The algebraic multiplicity of the zero symplectic eigenvalue  increases
at the point $V=V_c$. In fact from the hamiltonian property it follows that it increases up to 4
(rather than 3). 

Recalling the definition of the momentum integral \eqref{P} and
writing it in terms of the real and imaginary part of $\psi_s$, 
equation \eqref{F} becomes simply 
\[
\left. \frac{\partial P}{ \partial \chi} \right|_{\chi=\chi_0} =0. 
\]
This condition ensures that a two-parameter family of solutions $\psi_s(x-\theta; \chi)$,
existing at the velocity $V=V_0$, has a one-parameter subfamily $\psi_s(x-\theta; \chi_0)$ continuable 
to $V \neq V_c$ \cite{Baer}.

\section{Non-propagating solitons}
\label{Nonpropagating} 

\subsection{Simple solitons}

\begin{figure}
 \includegraphics[width =\linewidth]{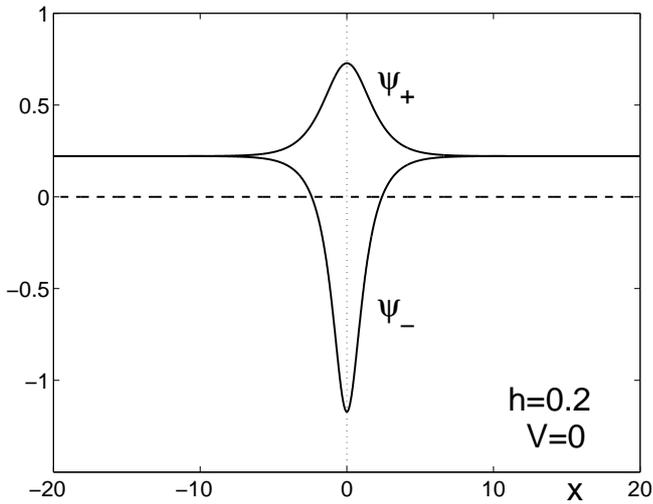}
\caption{\label{stat}
Stationary $\psi_+$ and $\psi_-$ solitons 
} 
\end{figure}

The ordinary differential equation \eqref{stationary} with $V=0$,
\begin{equation}
 \psi_{xx} + 2 |\psi|^2 \psi - \psi= -h,
\label{our5}
\end{equation}
 has two real-valued localised solutions, $\psi_+$ and $\psi_-$.  These are given by explicit formulas
 \cite{BZB}:
 \begin{equation}
\psi_\pm (x) = \psi_0
\left[ 
1+ \frac{2 \sinh^2 \beta}{1 \pm \cosh \beta \cosh (Ax)}
\right],\label{pm}
\end{equation}
where the parameter $\beta$ ($0\leq \beta <\infty$) is in one-to-one correspondence 
with the driving strength $h$:
\[
h = \frac{\sqrt{2} \cosh^2 \beta}{(1+ 2 \cosh^2 \beta)^{3/2}}.
\]
As 
$h$ increases from 0 to $\sqrt{2/27} \approx 0.2722$,
$\beta$ decreases from infinity to zero.
(Hence $0 \leq h \leq 0.2722$ is the domain of existence of the two solitons.)
The asymptotic value $\psi_0$ and inverse width $A$ are also expressible through $\beta$:
\[
\psi_0= \frac{1}{\sqrt{2}} \frac{1}{\sqrt{1+ 2 \cosh^2 \beta}},
\quad
A= \frac{\sqrt{2} \sinh \beta}{\sqrt{1+ 2 \cosh^2 \beta}}.
\]
(Note that the asymptotic value $\psi_0$ corresponds to the stable background,
denoted $\psi_1$ in Sec.\ref{Flat}.)

The stationary soliton $\psi_+$ has a positive eigenvalue in the spectrum
of the linearised operator \eqref{EV};  hence the $\psi_+$ is unstable for all $h$ for which it exists
\cite{BZB}. 
The spectrum of the stationary soliton $\psi_-$ with small $h$ includes two
discrete eigenvalues $\lambda_{1,2}=i \omega_{1,2}$, $\omega_{1,2}>0$ --- and their negative-imaginary counterparts. 
As $h$ grows to 0.07749, $\lambda_1$ and $\lambda_2$ approach each other, collide
and acquire real parts of the opposite sign. 
This is a hamiltonian Hopf bifurcation. For $h>0.07749$, the soliton $\psi_-$ is prone to the oscillatory instability
\cite{BZB}.

When a damping term is added to the equation, 
 the two stationary solitons $\psi_+$ and $\psi_-$ 
persist and can form a variety of multisoliton bound states, or complexes \cite{Malomed,Wabnitz,BSA,Kollmann}.
In the next subsection, we  show that {\it undamped\/} directly driven solitons
can also form stationary complexes.  Some of these complexes are bound so tightly that 
 the solution represents a single entity.
To distinguish these objects from the solitons
$\psi_+$ and $\psi_-$,  we will be referring to the $\psi_+$ and $\psi_-$
as the {\it simple\/} solitons.

\subsection{The twist solitons}

\begin{widetext}

\begin{figure}
 \includegraphics[width =0.49\linewidth]{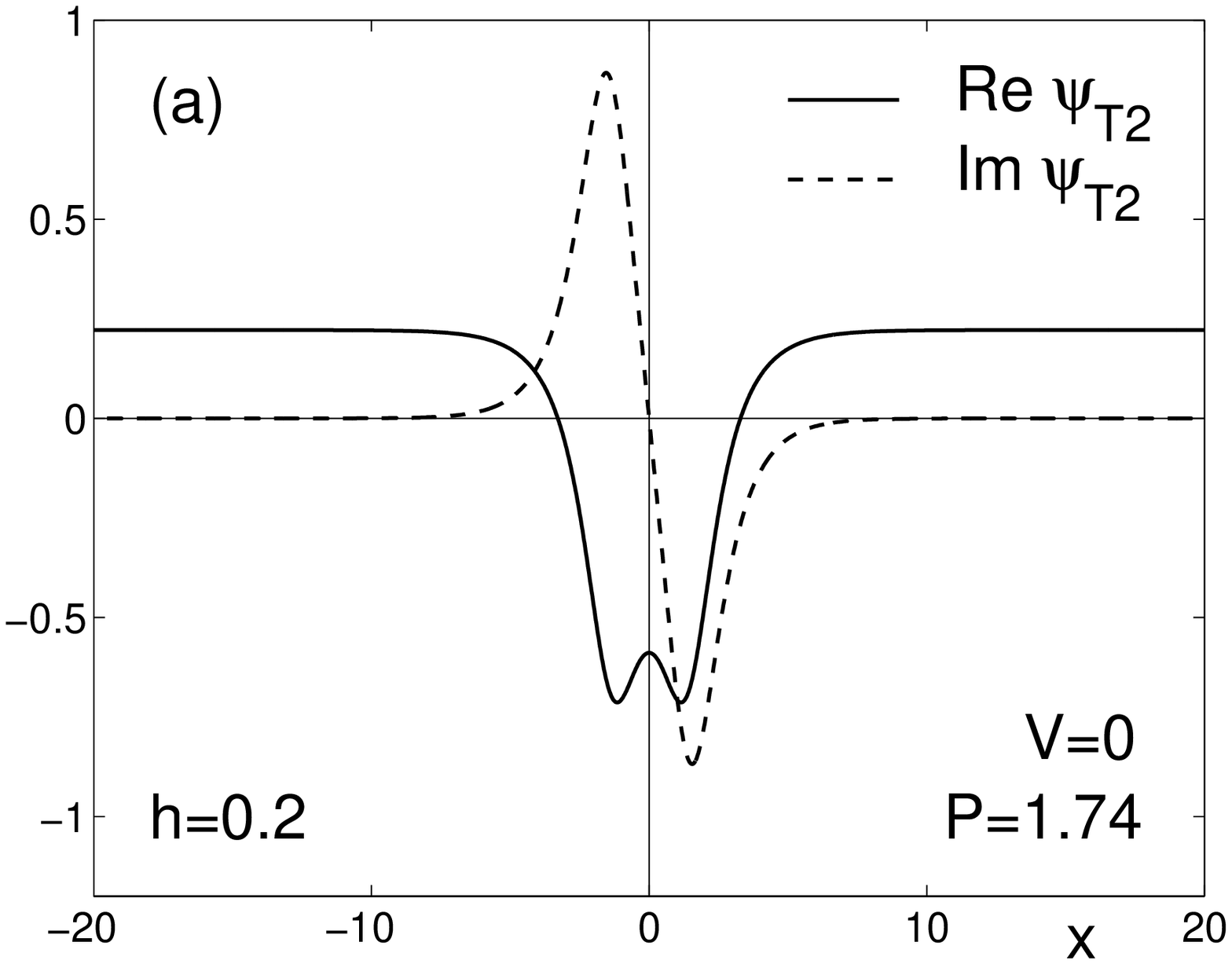}  
     \includegraphics[width =0.49\linewidth]{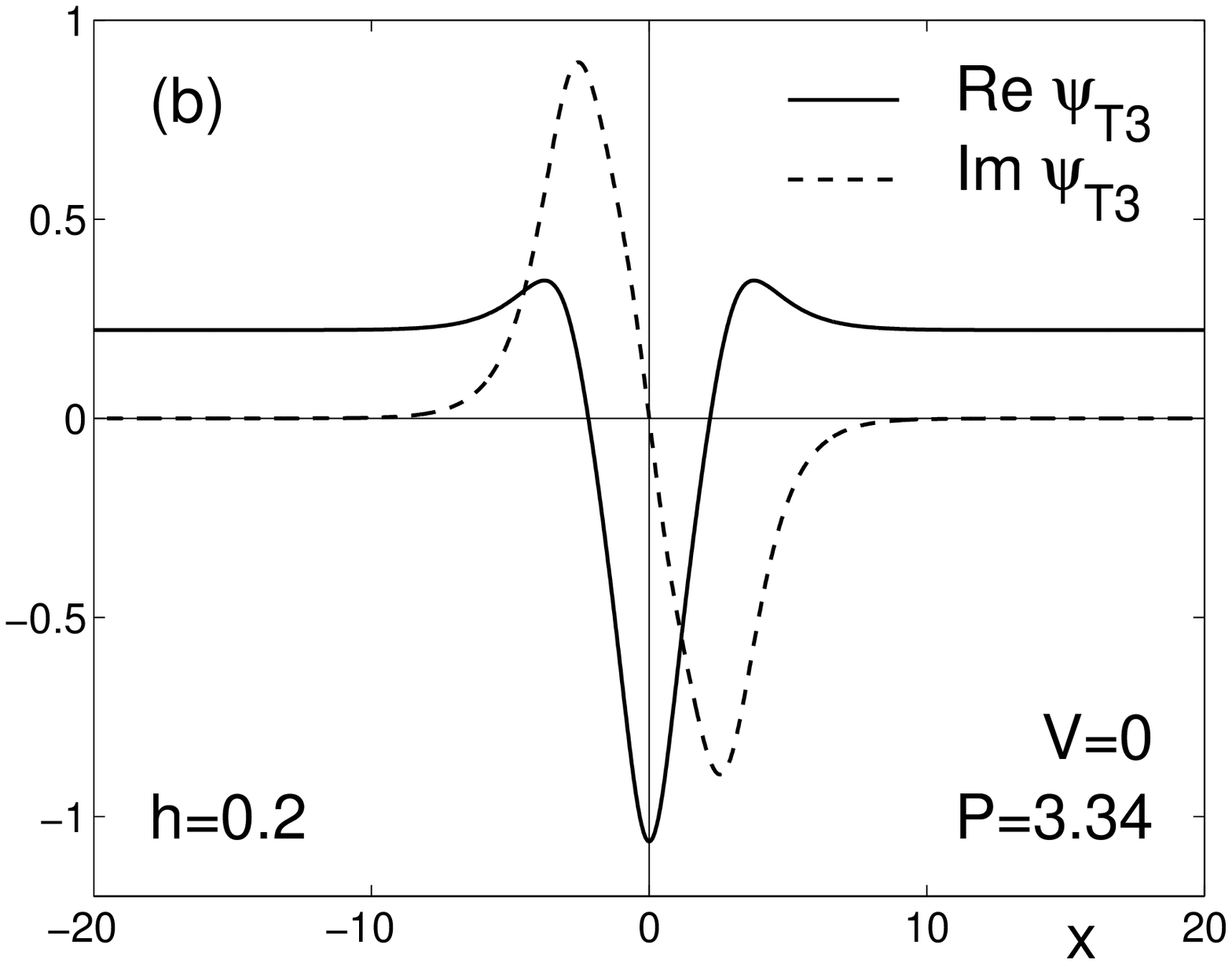}
      \includegraphics[width =0.49 \linewidth]{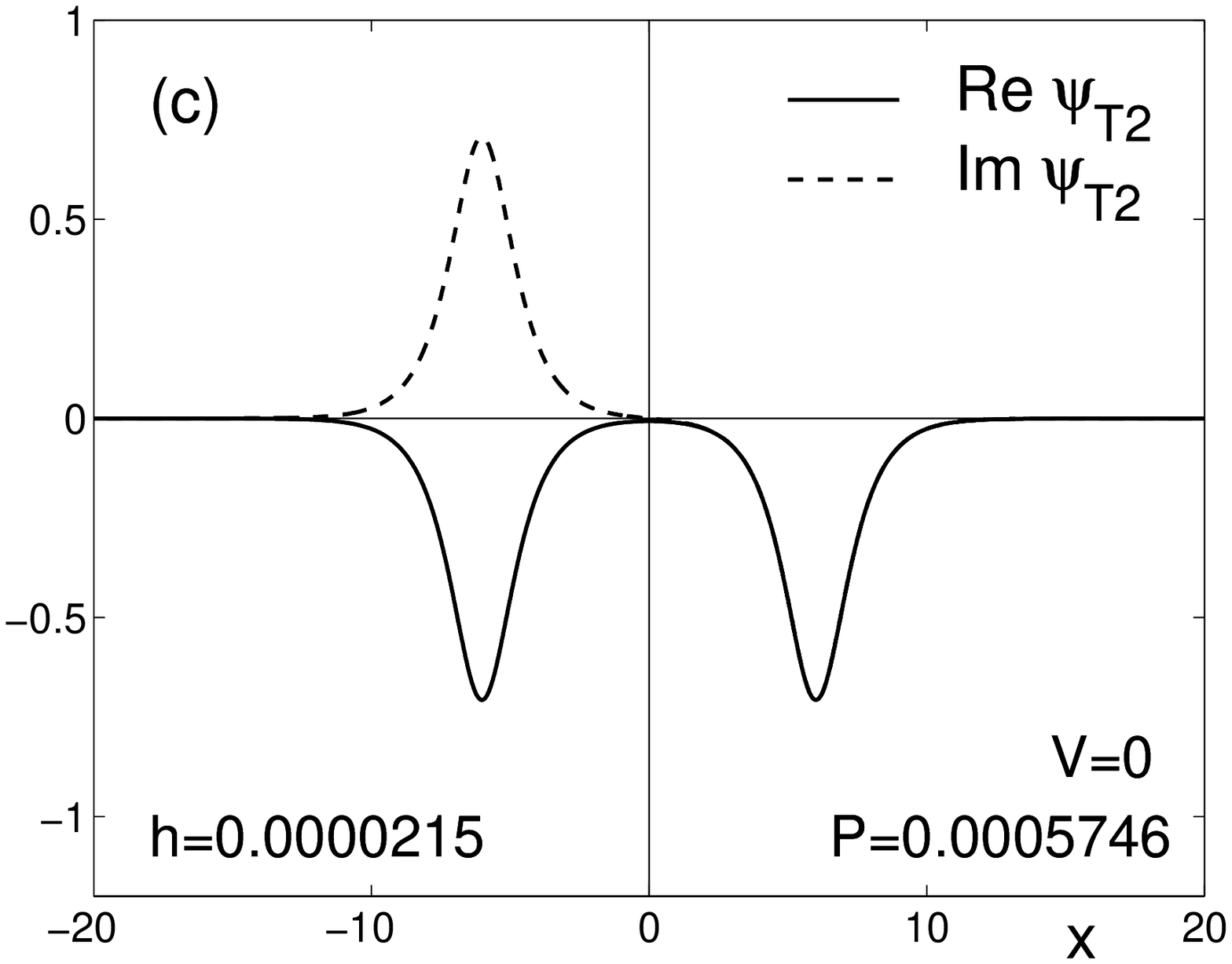}
        \includegraphics[width =0.49\linewidth]{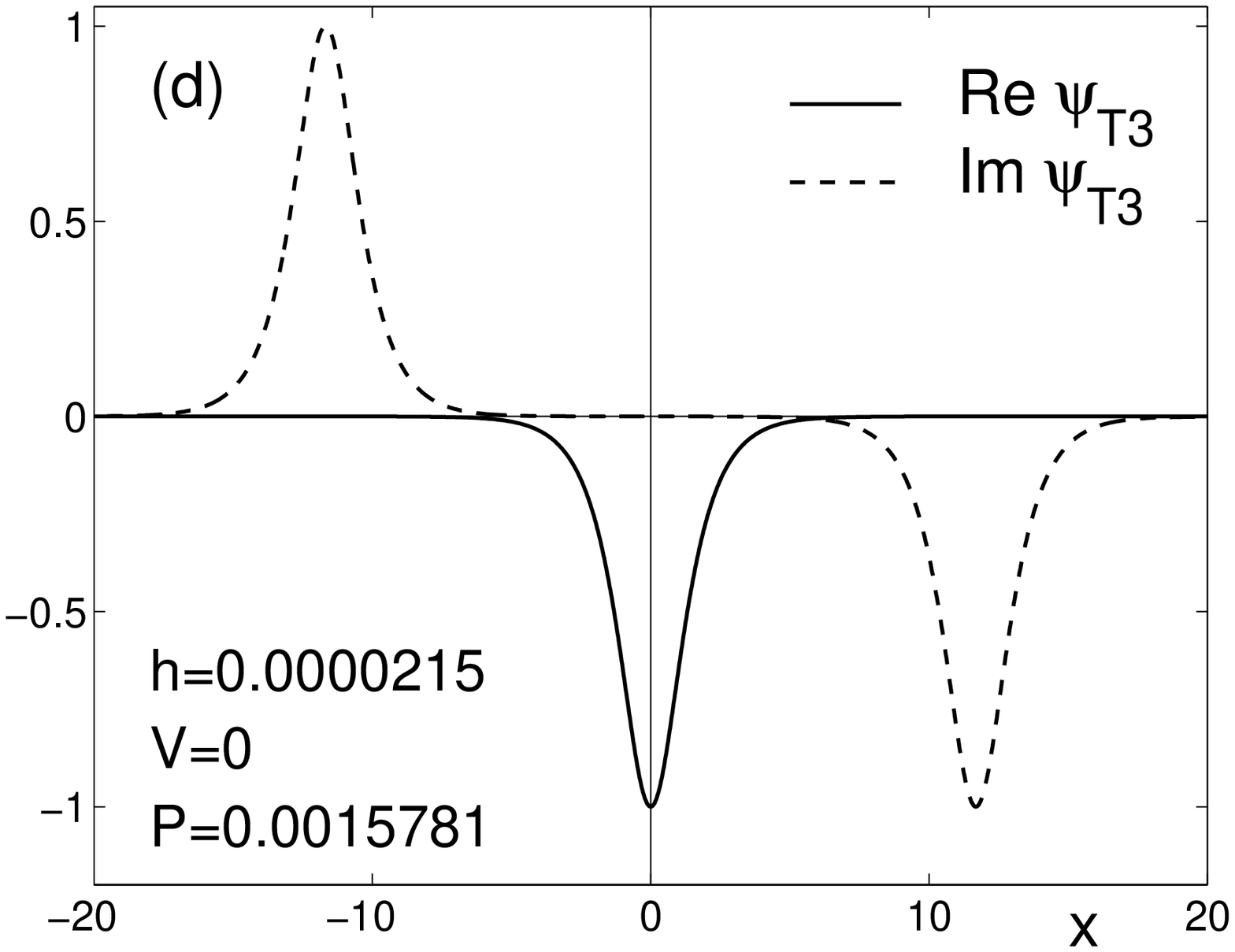}    
       \caption{\label{twi}
 (a,b): The two nonpropagating twist 
solutions for $h \sim 1$. (Here $h=0.2$).
(c,d): The corresponding quiescent solutions when continued to an exponentially small $h$.
(Here $h=2.15 \times 10^{-5}$).
All twist solutions shown in these figures are left-handed.
} 
\end{figure}
\end{widetext}

In addition to the two simple solitons expressible in elementary functions, 
 the stationary equation
\eqref{our5} has two localised solutions that cannot be constructed analytically. 
Unless  $h$ is extremely small, each of these two solutions has the form of a single entity
[Fig.\ref{twi}(a,b)] ---
a soliton whose phase does not stay constant but grows, monotonically, as $x$
changes from large negative to large positive values. 
When visualised in the three-dimensional $(x, \mathrm{Re}  \psi, \mathrm{Im} \psi)$-space, 
it looks like a  twisted ribbon (twisted by $360^\circ$); hence we will be calling these two 
solutions simply ``twists". For the reason that will become obvious 
in the paragraph following the next one, we denote the two solutions
$\psi_{T_2}$ and $\psi_{T_3}$, respectively. 

The twist solitons were previously encountered in the parametrically driven (undamped) 
nonlinear Schr\"odinger equation \cite{Baer}.  For each $h$, the parametrically driven twist is a member of a 
two-parameter family of stationary {\it two}-soliton solutions.
The first parameter is the overall translation  of the complex;
the second one is the separation distance between the two bound solitons. The twist corresponds to
a very small separation, where the two simple solitons bind to form a 
single entity. (The resulting object does not have even a slightest reminiscence of a 
two-soliton state; without knowing the whole family, the relation would hardly be possible to guess.) 

The two simple solitons, $\psi_+$ and $\psi_-$,  detach from the
$U(1)$-symmetric family of solitons
of the unperturbed nonlinear Schr\"odinger  at $h=0$ \cite{BZ_PhysicaD}.
The two twist solutions of \eqref{our5} also hail from the solitons of the unperturbed 
equation; however this time the relation is more complicated. 
Reducing $h$, the two solutions transform into complexes
of well-separated solitons
[Fig.\ref{twi}(c,d)].  Namely, one of the two  twist solutions becomes 
a complex of two solitons:
\[
\psi_{T2} \to    e^{3i \pi/4}           \sech(x+x_0)
+   e^{-3i \pi/4}  \sech(x-x_0),
\]
where $x_0 \to \infty$ as $h \to 0$. The other twist
continues to a complex of {\it three\/} unperturbed solitons:
\[
\psi_{T3} \to \mathrm{i} \,  \sech(x+x_0) -\sech x - \mathrm{i } \,  \sech(x-x_0),
\]
and again, the separation $x_0$ grows without bound as $h \to 0$. 
The ``full names" of the two twists, $\psi_{T2}$ and $\psi_{T3}$, were coined to reflect
this multisoliton ancestry.

Despite being quiescent, nonpropagating objects, the twists carry
nonzero momentum. Since equation \eqref{our5} is invariant 
under the space inversion, the twist soliton with momentum $P$ has a partner
with momentum $-P$ which is obtained by changing $x \to -x$. 
This transformation leaves the absolute value of $\psi(x)$ intact 
but changes the sign of the phase derivative, $(d/dx) \mathrm{arg} \, \psi(x)$. By analogy
with the right-hand rule of circular motion, the twist whose phase decreases as $x$ grows from $-\infty$
to $+\infty$ [that is, the trajectory on the $(\mathrm{Re} \, \psi, \mathrm{Im} \, \psi)$ phase plane
is traced clockwise], will be called right-handed. 
The twist with the increasing phase (i.e. with a trajectory traced counter-clockwise)
will be called left-handed.
One can readily verify that the left-handed twist has a positive
momentum, whereas the right-handedness  implies $P<0$. 

Consider some particular value of the driving strength, $h=h_0$.
Unlike the twist solution in the {\it parametrically\/} driven NLS, the {\it directly\/}
driven twist with $h=h_0$ is a member of a one-parameter
(rather than two-parameter)  family of solutions. 
(The only free parameter is the translation, $-\infty<\theta< \infty$, whereas
the intersoliton separation $\chi$ is fixed by $h$.)
This can be concluded from the fact that 
the corresponding operator $\mathcal{H}$ has only one, translational, 
zero eigenvalue. Had the twist been a member of a  family of solutions 
parametrised by two continuous parameters, say $\theta$ and $\chi$, the operator $\mathcal{H}$
would have had an additional zero eigenvalue with the eigenvector ${\vec \Psi}_\chi$.

Letting $\psi= x_1+ix_2$, the stationary equation \eqref{our5}
can be written  as a  classical mechanical system on the plane, with the Lagrangian
\[
L= \frac12 ({\dot x}_1^2+ {\dot x}_2^2) - \frac12 (x_1^2+ x_2^2)^2
+ \frac12 (x_1^2+ x_2^2) -h x_1.
\]
The existence of a one-parameter family of homoclinic orbits ${\vec x}={\vec x}_\chi(t)$,
where ${\vec x} \equiv(x_1,x_2)$,
would  imply that 
the above system has the second integral of motion,
in addition to the energy.
However, equation \eqref{our5} is known not to have
any additional conserved quantities \cite{Hietarinta}.

Finally, we need to comment on the stability of the two twist solutions.
When $h$ is equal to 0 and the two solutions represent a doublet and a triplet of
infinitely separated solitons of the unperturbed nonlinear Schr\"odinger, 
 the symplectic spectrum includes 8 and 12 zero eigenvalues, respectively.
 When $h$ is small nonzero, only two eigenvalues remain at the origin in each case.
 In addition, the spectrum of the $\psi_{T2}$ twist includes a complex quadruplet
 $\pm \lambda, \pm \lambda^*$ and a pair of opposite pure imaginary eigenvalues. As 
 $h$ is increased, the imaginary pair collides with another imaginary pair 
 emerging from the continuum, producing the second complex quadruplet.
 The spectrum of the $\psi_{T3}$ twist includes two complex quadruplets
 and a pair of pure imaginary eigenvalues; this arrangement remains in place for
 all $h$, from very small to $h=\sqrt{2/27}$.
 The bottom line is that both twist solutions are unstable for all $h$; the instability
 is always of the oscillatory type.

\section{Numerical continuation of simple
solitons}
\label{Numerical_Simple}

\subsection{The travelling $\psi_+$ soliton}

Travelling solitons are sought as 
solutions of  the ordinary differential equation \eqref{stationary}
under the boundary conditions $\psi_\xi \to 0$ as $|\xi| \to \infty$.

\begin{figure}
 \includegraphics[width =\linewidth]{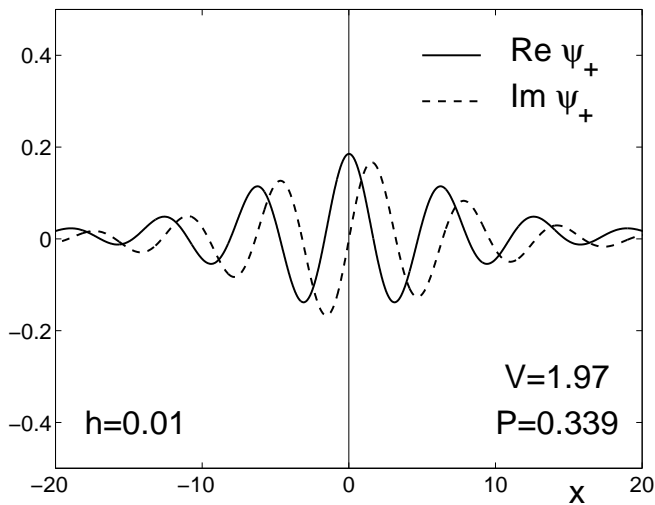}
\caption{\label{Vc}
As $V \to c$, the $\psi_+$ solitons (for all $h$) and $\psi_-$ solitons 
(for small $h$) approach 
linear waves with slowly decaying envelopes.
Shown is the $\psi_+$ solution with $V$ close to $c$.
(In this plot, $h=0.01$; the corresponding $c=1.9996$.)
The $\psi_-$ solutions with $V$ close to $c$ have a similar shape.}
\end{figure}

We begin with the continuation of the quiescent soliton $\psi_+$. 
For a sequence of  $h$ sampling the interval $(0, \sqrt{2/27})$, the branch 
starting at $\psi_+$ was path followed all the 
way to $V=c$, where $c$ is given by Eq.\eqref{c}. As $V$ increases, 
the amplitude of the solution  decreases while
the width grows. A typical solution with $V$ close to $c$ is shown  in Fig.\ref{Vc}.
As $V \to c$,
the momentum $P$ tends to zero.

\begin{widetext}

\begin{figure}
\includegraphics[width =0.49\linewidth]{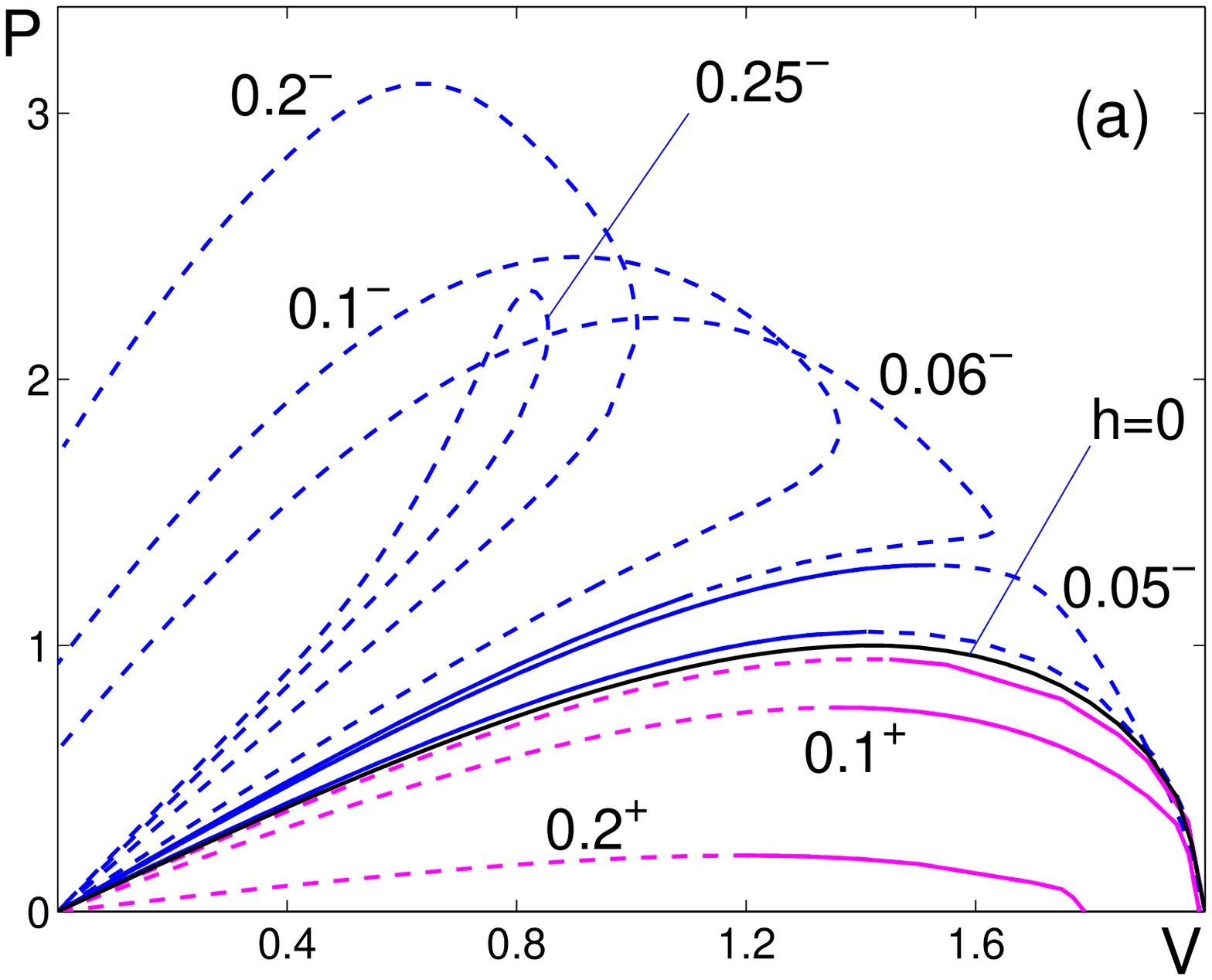}
\hspace{0.5mm}
\includegraphics[width =0.49\linewidth]{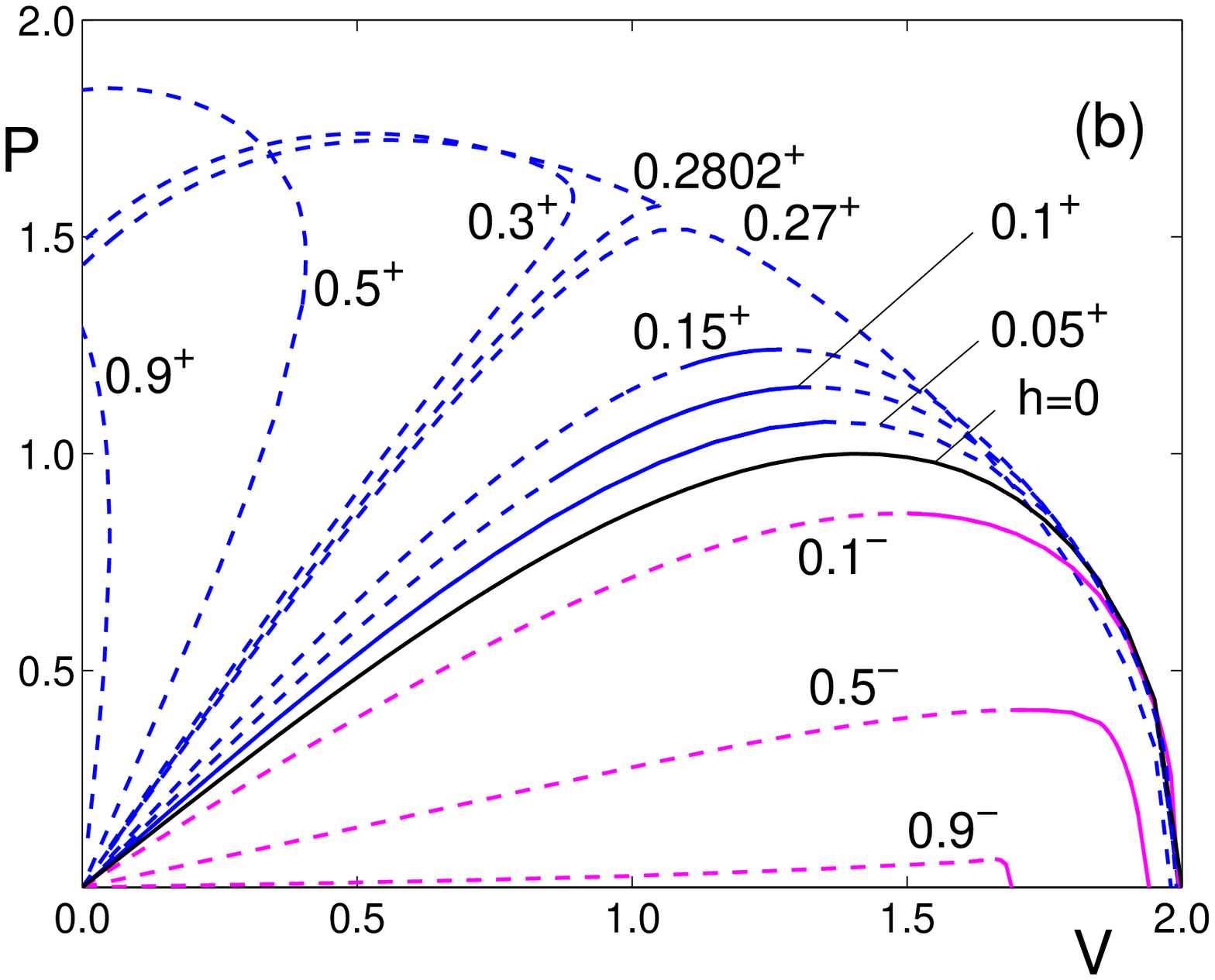}
\caption{\label{PV} (a) The momentum  of the $\psi_+$, $\psi_-$ 
solitons continued to positive velocities. Decimal fractions attached to branches label the corresponding values of $h$, with the superscripts 
$+$ and  $-$ indicating the $\psi_+$ and $\psi_-$ solitons. (For example, $0.1^+$ 
marks the branch emanating from  the stationary $\psi_+$ soliton with $h=0.1$.) 
The only two branches that have not been labelled
 are the $\psi_+$ and $\psi_-$ branches
with $h=0.01$; the curve just above $h=0$ is $0.01^-$ and the curve just below
the $h=0$ branch is $0.01^+$.
Solid curves mark stable and dashed ones unstable branches.
 (b)  The corresponding $P(V)$ diagram for the travelling parametrically driven solitons 
 from \cite{Baer}.}
\end{figure}
\end{widetext}

The resulting $P(V)$ diagram is shown in Fig.\ref{PV}(a). 
For each $h$, the unstable stationary $\psi_+$ soliton  remains unstable
when travelling sufficiently slow. The instability is due to a real eigenvalue $\lambda>0$ of
the linearised operator \eqref{EV}.

 As $V$ grows, the unstable eigenvalue moves towards the 
origin along the real axis. 
Eventually, as the momentum $P$ reaches its maximum, the positive eigenvalue $\lambda$
collides with its opposite partner $\lambda'=-\lambda$, after which 
both real eigenvalues move onto the imaginary axis and the soliton acquires stability.
The soliton remains stable all the way from the point
 $V_c$, where
the momentum is maximum, to the value $V=c$ where $P=0$ and the soliton ceases to exist.

The resulting $P(V)$ dependence shows a remarkable similarity to the $P(V)$ diagram \cite{Baer}
 for the {\it parametrically\/} driven nonlinear Schr\"odinger,
 \begin{equation}
i \psi_t -iV \psi_\xi + \psi_{\xi \xi} + 2 |\psi|^2 \psi - \psi= h \psi^*.
\label{PDNLS}
\end{equation} 
 The ``parametrically driven" diagram is reproduced in Fig.\ref{PV}(b) for the sake of comparison.  One should keep in mind here
that the notation used for the parametrically driven solitons is opposite to the notation employed 
in the externally driven situation.
Thus, the parametrically driven stationary ($V=0$)  soliton with a positive 
symplectic eigenvalue in its spectrum 
 is denoted $\psi_-$  (and not $\psi_+$ as its externally driven counterpart). 
On the other hand, 
 the parametrically driven stationary soliton denoted $\psi_+$ 
is stable for sufficiently small $h$ (like the externally driven soliton $\psi_-$).
For this reason, the  objects
featuring $P(V)$ diagrams similar to those of our externally driven  solitons $\psi_+$, are the parametrically driven solitons $\psi_-$.

\subsection{The travelling $\psi_-$ soliton; $h <0.06$}

In the case of the $\psi_-$ solitons, there  are two characteristic scenarios. 
When $h$ lies between $0$ and $0.06$, the soliton $\psi_-$ exists for all $V$ between $0$ and $c$.
 As $V$ is increased from zero, the momentum $P$ grows from $P=0$ and
 reaches its maximum at some point $V_c$, $0<V_c<c$.
As $V$ is changed from $V_c$ to $c$, the momentum 
decays to zero [see Fig.\ref{PV}(a)].  On the other hand, 
 when $h$ equals $0.06$ or lies above this value, the curve $P(V)$ does not
exhibit a point of maximum.

Consider, first, the case $h<0.06$. 
The transformation scenario here is similar to the case of the soliton $\psi_+$;
see Fig.\ref{PV}. What makes 
the bifurcation curves for the $\psi_+$ and $\psi_-$ solitons 
different, is the stability properties of the two solutions. 
Unlike the $\psi_+$ solution, 
the {\it stationary\/}  $\psi_-$ soliton with  $h \leq 0.07749$ is stable and its stability persists when it is continued to
small nonzero velocities. As $V$ grows to the value $V_c$ where the momentum reaches its maximum, 
two opposite  pure imaginary eigenvalues collide at the origin on the $(\mathrm{Re} \lambda, \mathrm{Im} \lambda )$
plane and cross to the positive and negative real axis, respectively.
 For the driving strengths
$h \leq 0.055$, this implies the loss of stability. 

As for the interval $0.0551 \leq h \leq 0.06$, here the instability sets in earlier, as  $V$ reaches some $V=V_0$
(where $V_0<V_c$).  At the point $V=V_0$, 
two pairs of pure imaginary eigenvalues collide and produce a quadruplet of complex eigenvalues $\pm \lambda, \pm \lambda^*$. (Here $\lambda$ has a small real and finite imaginary part.)
This is a point of the  hamiltonian Hopf bifurcation, associated with the oscillatory 
instability  \cite{ABP,Baer}. 
As $V$ is  increased to $V_1$
(where $V_0<V_1<V_c$), two  pairs of complex-conjugate $\lambda$
converge on the real axis, becoming two positive 
($\lambda_1=\lambda_2>0$) and two negative ($-\lambda_1=-\lambda_2$)
eigenvalues. Finally, when  $V$ crosses through $V_c$, the eigenvalues $\lambda_1$ and $-\lambda_1$ 
move on to the imaginary axis. The soliton does not restabilise at this point though;
the real pair $\pm \lambda_2$ persists for all $V \geq V_c$.

The bifurcation values $V_0$  and $V_1$ are, naturally, functions of $h$. 
The value $V_0$ decreases 
(and $V_1$ increases) as $h$ is increased from $0.0551$.  Eventually,
when $h$ reaches $0.07749$, $V_0$ reaches zero. It is interesting to
note that there is a gap between $V_1$ and $V_c$ for all $h$.
Therefore the oscillatory and nonoscillatory instability coexist for
no  $V$; for smaller $V$ ($V_0 < V <V_1$) the instability is 
oscillatory whereas for larger $V$ ($V>V_1$) the instability has a
monotonic growth.

Finally, it is appropriate to
mention here that the bifurcation curve for the $\psi_-$ solitons with small $h<0.06$ 
has the same form as the $P(V)$ dependence
for the small-$h$ {\it parametrically\/} 
driven solitons  (more specifically, parametrically
driven $\psi$-{\it plus\/}  solitons) --- see Fig.\ref{PV}(b).

\subsection{The travelling $\psi_-$ soliton; $h \geq 0.06$}

The $P(V)$ graphs for $h \geq 0.06$ are qualitatively different
from the small-$h$ bifurcation curves. For these larger $h$, the 
bifurcation curve emanating from the origin on the $(V,P)$-plane
turns back at some $V=V_\textrm{max}$,  with the derivative $\partial P/ \partial V$ remaining strictly positive
for all $V \leq V_\textrm{max}$.

\begin{figure}
\includegraphics[width =\linewidth]{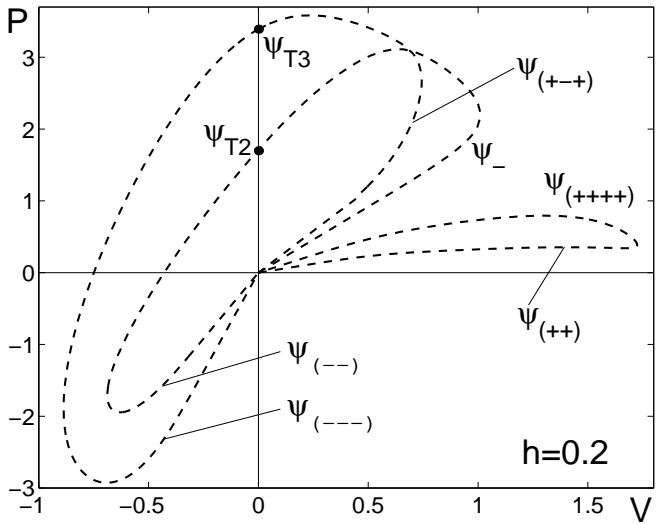}
\caption{\label{PV2} 
The  full $P(V)$ bifurcation diagram for the $\psi_-$ soliton with 
$h=0.2$.
Also shown is the continuation of the $T3$ twist and 
the $\psi_{(++)}$  branch.
 More solution branches can be
obtained by the reflection $V \to -V$, $P \to -P$.
All branches shown in this figure correspond to unstable solutions.
}
\end{figure}

For $h$ in the interval $0.06 \leq h <0.25$, the $P(V)$ curve crosses the
$P$-axis [Fig.\ref{PV}(a), Fig.\ref{PV2}]. 
The solution arising at the point $V=0$ is nothing but the $\psi_{T2}$ twist soliton,
shown in Fig.\ref{twi}(a).

As we continue this branch to the $V<0$-region, the twist transforms into a complex of
two well-separated $\psi_-$ solitons. 
The $P(V)$ curve makes one more turn
 and eventually returns to the origin on the $(V,P)$-plane (Fig.\ref{PV2}). 
As $V$ and $P$ approach zero, the distance between 
the solitons in the complex tends to infinity.

An interesting scenario  arises when $h$ is greater or equal than $0.25$.
Here, as $V$ grows from zero,  the soliton $\psi_-$ gradually transforms into a 
three-soliton
complex $\psi_{(+-+)}$. The branch 
turns back towards $V=0$ but does not cross the $P$-axis. Instead of continuing to negative $V$, 
the branch reapproaches the origin in the $(V,P)$ plane, remaining
in the positive $(V,P)$ quadrant
at all times. The ingoing path is almost 
coincident with the outgoing trajectory; as a result, the branch forms a lasso-looking loop [Fig.\ref{PV}(a)].

Turning to the stability properties of
solutions along the branch continued from $\psi_-$, we start with a 
short interval $0.06 \leq h \leq 0.07749$. 
The movements of the stability eigenvalues along the section of the curve emanating from 
the origin on the $(V,P)$ plane, 
are similar to the interval $0.055 < h <0.06$ 
that we discussed in the previous paragraph. The stationary $\psi_-$ soliton is stable
and stability persists for small $V$. As $V$ reaches a certain $V_0>0$, a quadruplet of complex eigenvalues is 
born and oscillatory instability sets in. 
Subsequently two pairs of the complex eigenvalues converge on the real axis, 
dissociate, then recombine
and diverge to the complex plane again;  a pair of opposite pure imaginary eigenvalues 
moves to the real axis and back --- however, despite all this activity on the complex plane, 
 the soliton solution never regains its stability.

For larger $h$, $h>0.07749$, the stationary $\psi_-$ soliton is unstable,
with a complex quadruplet in its spectrum.  As we continue in $V$, two pairs of opposite 
pure imaginary eigenvalues move on to the real axis, one after another.
For  $h \geq 0.25$, the resulting arrangement
(two pairs of opposite real eigenvalues and a complex quadruplet)
persists until the branch reaches the origin on the $(V,P)$ plane. 
On the other hand,  when $h$ lies in the interval $0.07749 <h<0.25$,
the four real eigenvalues collide, pairwise, producing the second complex quadruplet
at some point on the curve before it crosses the $P$ axis in Figs.\ref{PV}(a) and \ref{PV2}. 
Two complex quadruplets persist in the spectrum as we continue the curve
further.   Thus the unstable stationary 
soliton   $\psi_-$   with $h>0.07749$,  remains unstable for all $V$.

\section{Numerical continuation of the twist soliton}
\label{Numerical_Twist}

When $0.06 \leq h <0.25$, the branch resulting from the continuation of
the stationary $\psi_-$ soliton turns back and crosses the $P$-axis; the point of crossing 
corresponds to the $T2$ twist solution. On the other hand, when $h$ lies outside the 
$(0.06, 0.25)$ interval, the stationary $T2$ twist is disconnected from the 
stationary $\psi_-$ soliton and can be used as a starting point for a new,
 independent, branch. Another new branch is seeded by the $T3$ solution. 
 
These additional branches of travelling solitons are traced in this section.

\subsection{Travelling twist $T2$ ($h < 0.06$)}

We start with the situation of small $h$: $h < 0.06$, and consider the 
$T2$ solution first.

\begin{widetext}

\begin{figure}
\includegraphics[width =0.49 \linewidth]{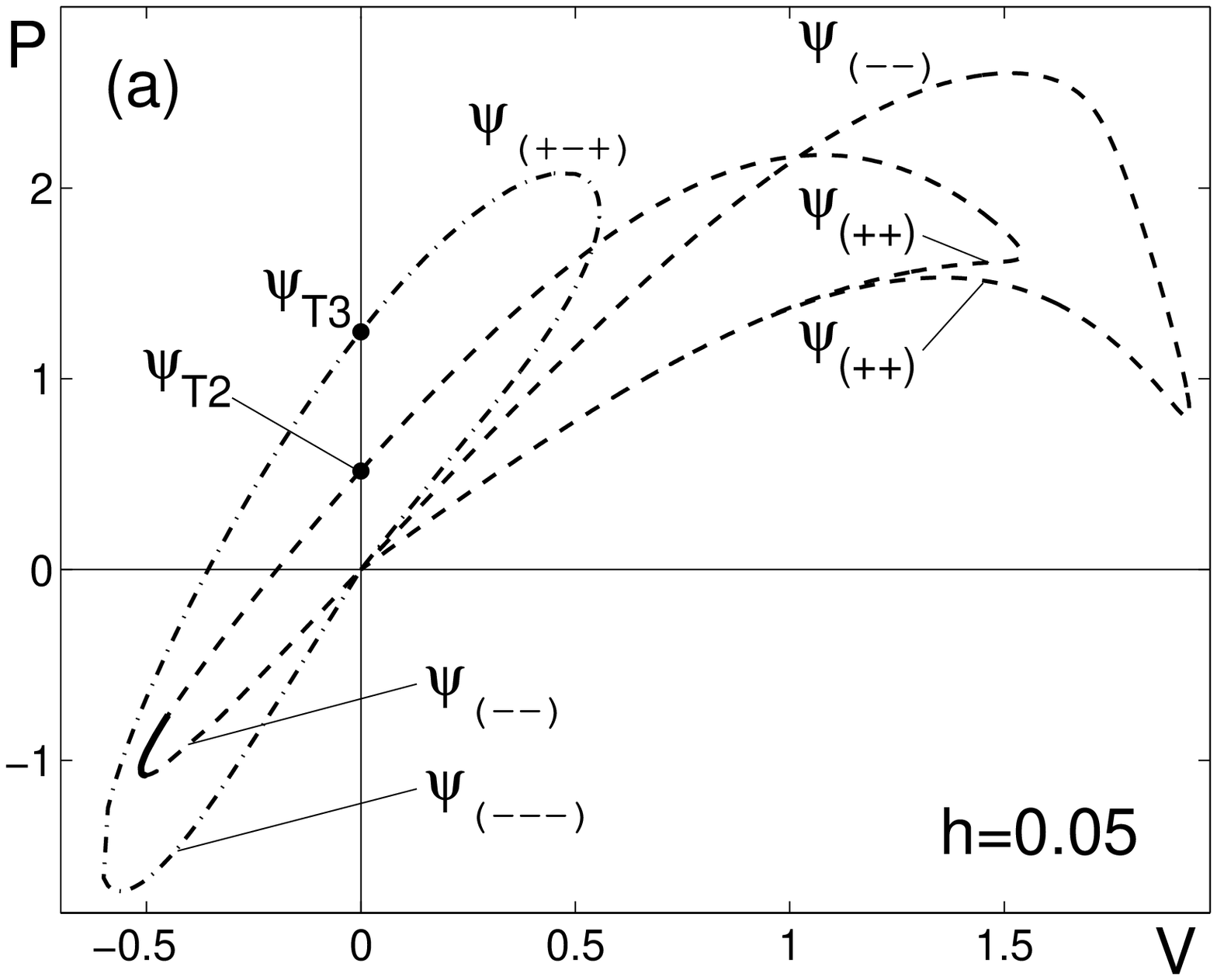}
\includegraphics[width =0.49 \linewidth]{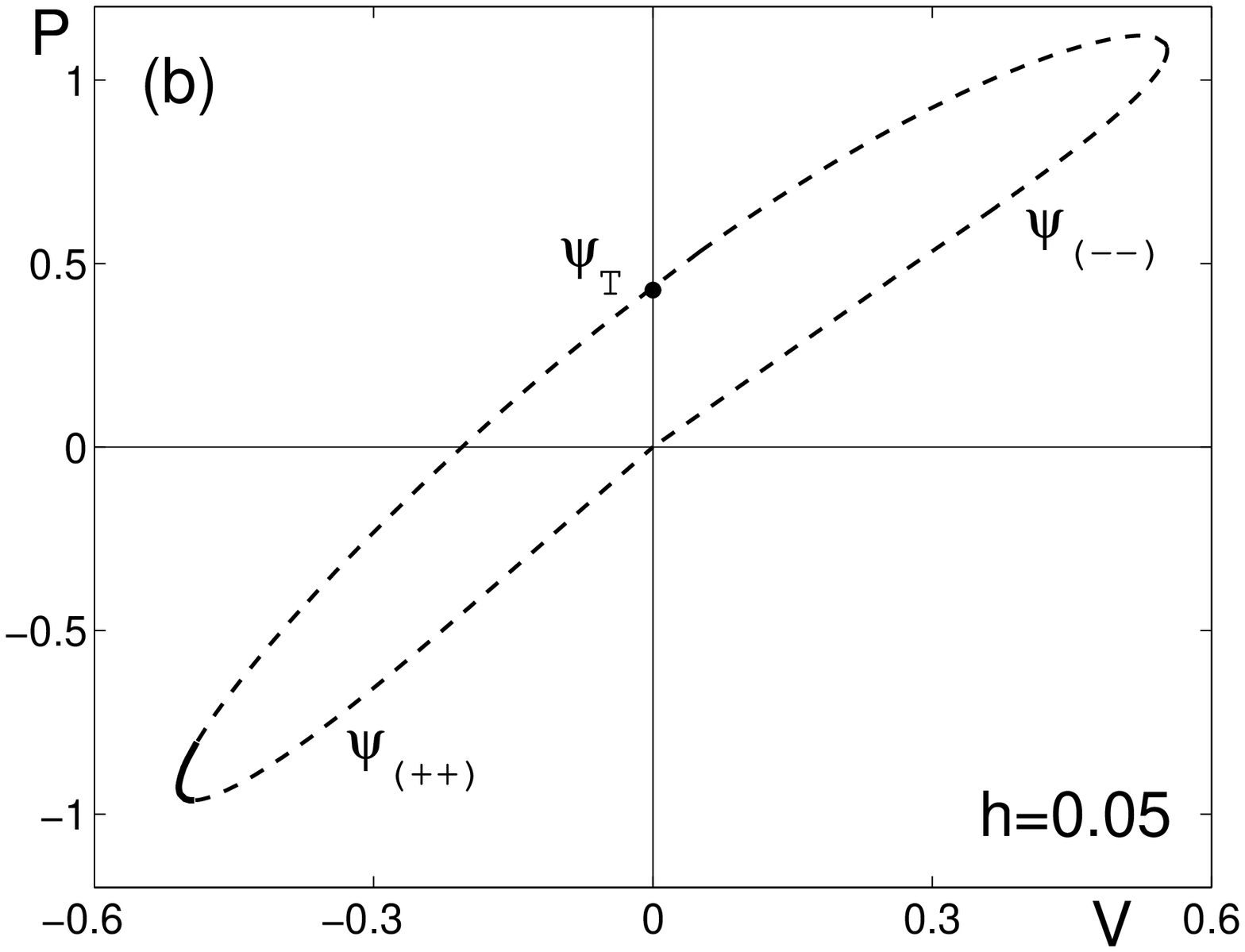} 
 \caption{\label{h05_complex}  (a) 
Continuation of the $T2$ and $T3$ twist solutions in the case of small $h$.
Also shown is the two-soliton branch which connects the origin to itself without
intersection the vertical axis. 
(b) The continuation of the twist soliton in the case of the 
parametrically driven NLS equation (adapted from \cite{Baer}).
 More solution branches can be
obtained by the reflection $V \to -V$, $P \to -P$ both in (a) and (b). 
}
\end{figure}
\end{widetext}

When the stationary $\psi_{T2}$ twist is path followed to positive $V$, it transforms into a $\psi_{(++)}$ complex.
At some point, the $P(V)$ curve makes a U-turn [Fig. \ref{h05_complex}(a)] and 
connects to the origin on the $(V,P)$ plane.
The entire positive-$V$ branch is unstable. The stationary twist has a complex quadruplet in its spectrum;
as the curve is continued beyond the turning point, the complex eigenvalues converge, pairwise,
on the positive and negative real axis. In addition,
a pair of opposite pure imaginary eigenvalues moves onto the real axis
as $V$ passes through the point of maximum of the
momentum in Fig.\ref{h05_complex}(a).

As the curve approaches the origin, the distance 
between the two  solitons in the complex 
increases and becomes infinite when $V=P=0$.
The spectrum becomes the spectrum of two infinitely separated $\psi_+$ solitons,
i.e. it includes two positive eigenvalues $\lambda_1 \approx \lambda_2$;
their negative counterparts $-\lambda_1 \approx -\lambda_2$;
and four eigenvalues near the origin.

Continuing the $T2$ twist in the negative-$V$ direction, it transforms into a complex
of two $\psi_-$ solitons. At some point along the curve, a quadruplet of complex eigenvalues
converges on the imaginary axis 
 and the complex
stabilises. 
(For the value $h=0.05$ which was used to produce Fig.\ref{h05_complex}(a), the stabilisation occurs at the point $V=-0.45$.)
Continuing to larger negative $V$, the branch turns back;
shortly after that (at $V=-0.503$ for $h=0.05$) the momentum reaches its minimum.
Two opposite imaginary eigenvalues collide at this point and move onto the real axis;
the solution loses its stability.

When continued beyond the turning point and the point of minimum
of momentum, 
 the curve connects to the origin on the $(V,P)$ plane (Fig. \ref{h05_complex}(a)).
 As $V,P \to 0$,
the distance between the two $\psi_-$ solitons grows without bound.
The two opposite real eigenvalues decay in absolute value but remain in the spectrum 
all the way to $V=0$.

It is interesting to  note a similarity between the bifurcation diagram 
resulting from the continuation of the small-$h$ $T2$ twist in the externally driven NLS
[Fig.\ref{h05_complex} (a)] and the corresponding diagram in the parametrically driven case.
The latter is reproduced, for convenience of comparison, in Fig.\ref{h05_complex} (b).
In both cases the continuation of the twist solution to negative velocities 
gives rise to a stable complex of two stable solitons.

\subsection{Travelling twist $T3$, $h<0.25$}

Figs.\ref{h05_complex}(a) 
and \ref{PV2}
also show the continuation of the $T3$ twist soliton. 
The bifurcation diagrams obtained for
$h<0.06$ and $0.06 \leq h <0.25$ are qualitatively similar.

Continuing the stationary $T3$
to positive velocities, the solution transforms into a $\psi_{(+-+)}$ complex.
If we, instead, continue to negative velocities, the twist transforms into a triplet of 
$\psi_-$ solitons. Both $V>0$ and $V<0$ parts of the curve turn 
and connect to 
the origin on the $(V,P)$ plane. As $V$ and $P$ approach the origin
on either side, the distance between the three solitons bound in the complex grows without limit.

The stationary $T3$ has two complex quadruplets in its spectrum;
depending on $h$, both or one of these converge on the real axis
as we continue it to $V>0$ and $V<0$.
Two opposite  eigenvalues cross through $\lambda=0$ at the 
extrema of $P(V)$. Finally,
as $V$ and $P$ approach the origin, the spectrum transforms into the union of spectra of three separate solitons.

\subsection{Travelling twists $T2$ and $T3$,  $h \geq 0.25$}

\begin{widetext}

\begin{figure}
 \includegraphics[width =0.49\linewidth]{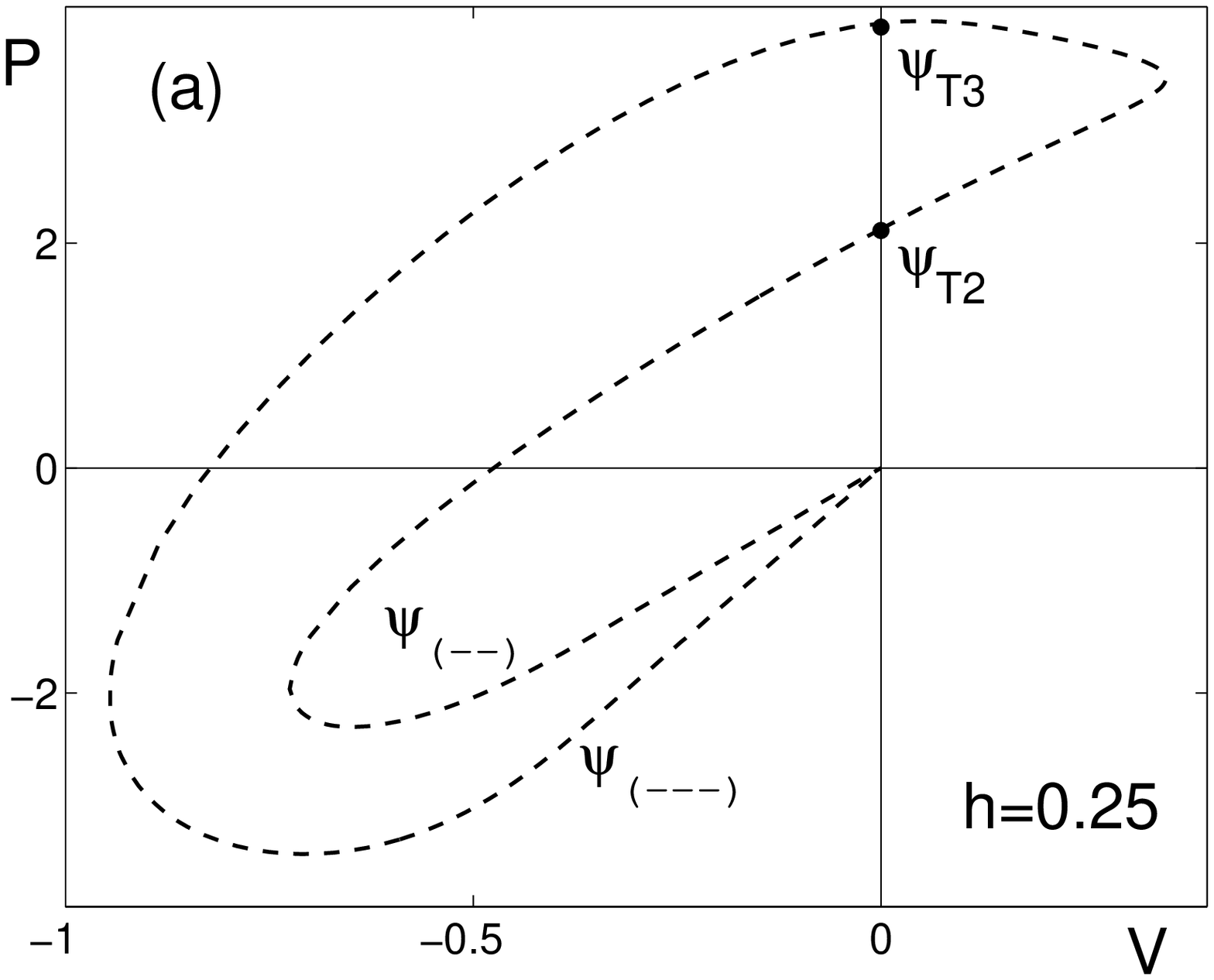}
 \includegraphics[width =0.49\linewidth]{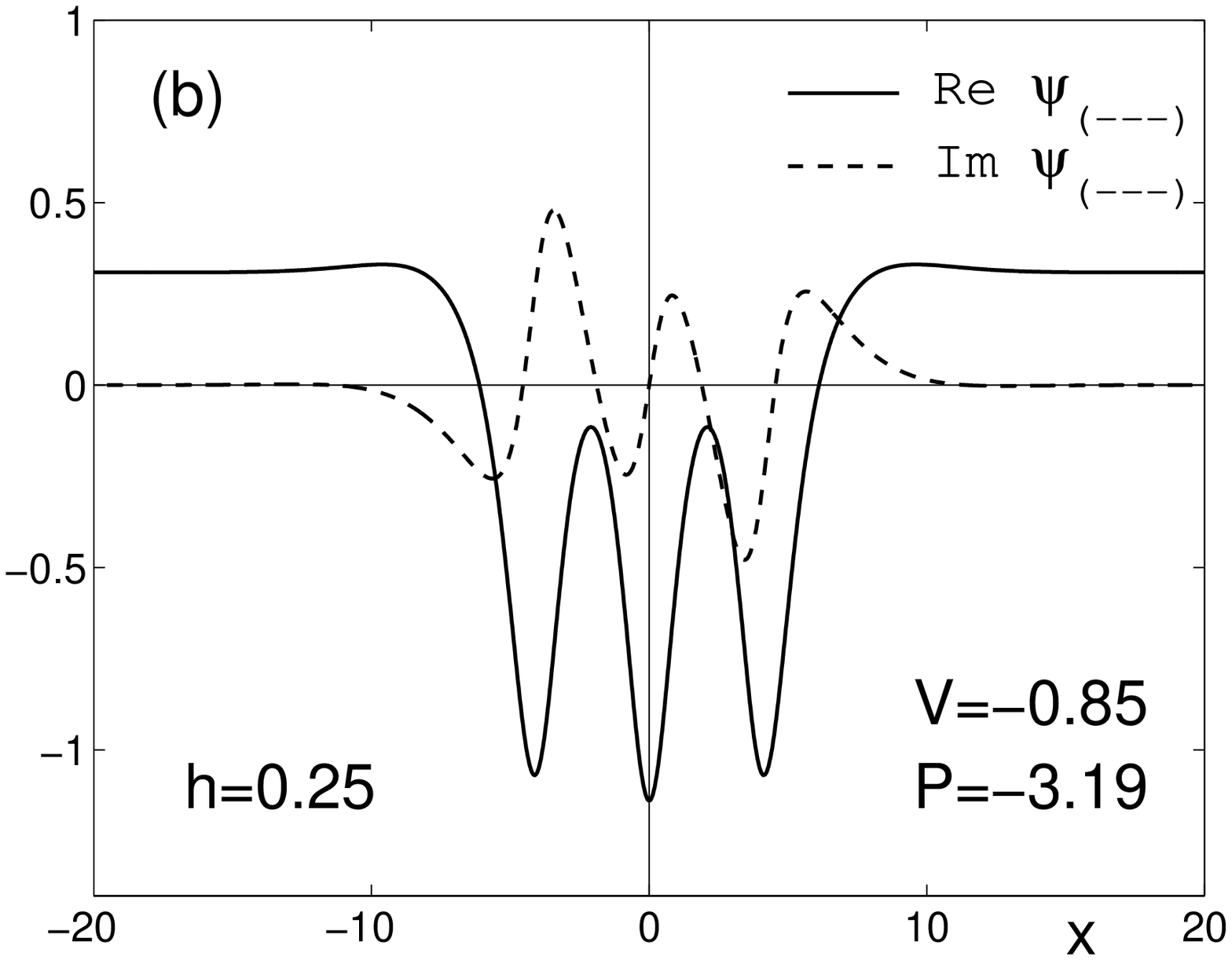} 
    \caption{\label{twist_25} (a) The $P(V)$ curve resulting from the 
    continuation of the twist for $h=0.25$. The starting point of the continuation is
    marked by an open circle. All branches shown in this figure are unstable.
       (b) A $\psi_{(---)}$  solution on the lower branch in (a). Here $V=-0.85$, $P=-3.2$. 
    In (b), the solid lines show the real and dashed imaginary part.
} 
\end{figure}
\end{widetext}

Another parameter region where  the continuation of the 
$\psi_-$ does not cross the $P$-axis, is $h \geq 0.25$. 
  The result of the continuation of the two
  twist solutions is shown in Fig.\ref{twist_25}(a). 
 The continuation of $T2$ to the negative velocities proceeds according to scenario
 similar to $h=0.2$ and $h=0.05$: the twist transforms into a complex of two
 solitons $\psi_-$. At some negative $V$ the curve turns back and  connects to the origin on the $(V,P)$ plane,
 with the distance between the two solitons bound in the complex increasing without bound.
 The eigenvalues evolve accordingly:  two complex quadruplets in the spectrum of the stationary $T2$ 
 persist for all $V<0$, supplemented by a pair of real eigenvalues which arrive from the
 imaginary axis at the point of minimum of $P(V)$. As $V,P \to 0$, the discrete spectrum 
 becomes the union of the eigenvalues of two simple solitons. 
 
 The continuation of $T2$ to positive $V$ produces a less expected outcome.
 Instead of turning clockwise and connecting to the origin
 as in Fig.\ref{h05_complex}(a), the curve turns counterclockwise
 and crosses through the $P$-axis once again. The solution arising at 
 the point $V=0$
 is nothing but the twist $T3$. 
 Two complex quadruplets in the spectrum of $T2$ persist as
 it is continued to $T3$.  
 
 The subsequent continuation produces a hook-shaped curve similar to the
 curve described in the previous paragraph and leading to the origin on the
 $(V,P)$-plane. The corresponding solution is a complex of three $\psi_-$ solitons,
 shown in Fig.\ref{twist_25}(b).  The third complex quadruplet 
 emerges at some $V$ before the turning point, and a pair of opposite 
 real eigenvalues arrives from the imaginary axis at the point of minimum of the momentum.
 As $V,P \to 0$, the distance between the solitons grows to infinity
 and the spectrum approaches the union of the eigenvalues of three 
 separate solitons $\psi_-$.

\subsection{Other branches}

It is appropriate to note that there are branches which do not originate on any
of the four stationary solutions listed above ($\psi_{\pm}$, $\psi_{T2}$ or $\psi_{T3}$).
The simplest of these emerge from the origin
on the $(V,P)$ plane as bound states of simple solitons with large separation.
One branch of this sort arises for $h \geq 0.06$ (Fig. \ref{PV2}).
It emerges from the origin as the $\psi_{(++)}$ and returns as the $\psi_{(++++)}$ complex.
The entire branch is unstable.

Next, 
unlike in the parametrically driven NLS, the same pair of {\it externally\/} driven travelling solitons
may bind at various distances. In particular, when $h$ is smaller than 0.06,
there is more than one bound state of
two $\psi_+$ solitons and more than one complex of two $\psi$-minuses.
Fig.\ref{h05_complex}(a) shows a branch $\psi_{(--)}$ that emerges from the 
origin in the first quadrant of the $(V,P)$ plane, describes a loop
and re-enters the origin --- this time as 
 a $\psi_{(++)}$ branch. Note that
 for small $V$ and $P$,  the re-entering $\psi_{(++)}$ branch is indistinguishable from 
 the other $\psi_{(++)}$ branch --- the one that continues from the twist solution. 
 (In a similar way,  the $V \to -V$, $P \to -P$ reflection of the $\psi_{(--)}$ branch
 overlaps with the small-$V,P$ section of  the $\psi_{(--)}$ branch arriving from the twist.)
 All solutions constituting this branch are unstable.

\section{Concluding remarks}
\label{Conclusions}

\begin{widetext}

\begin{figure}
\includegraphics[width =0.49\linewidth]{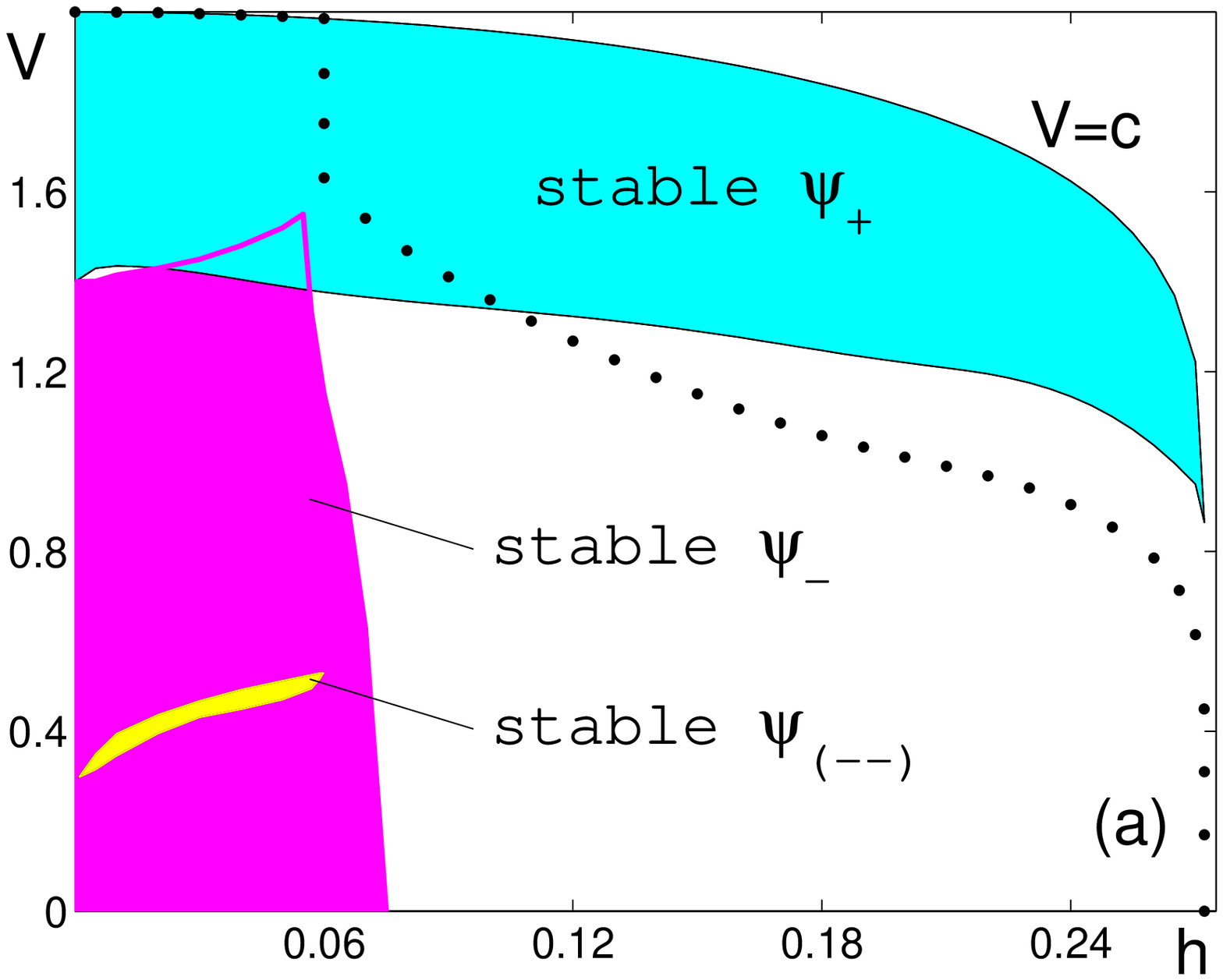}
\includegraphics[width =0.49\linewidth]{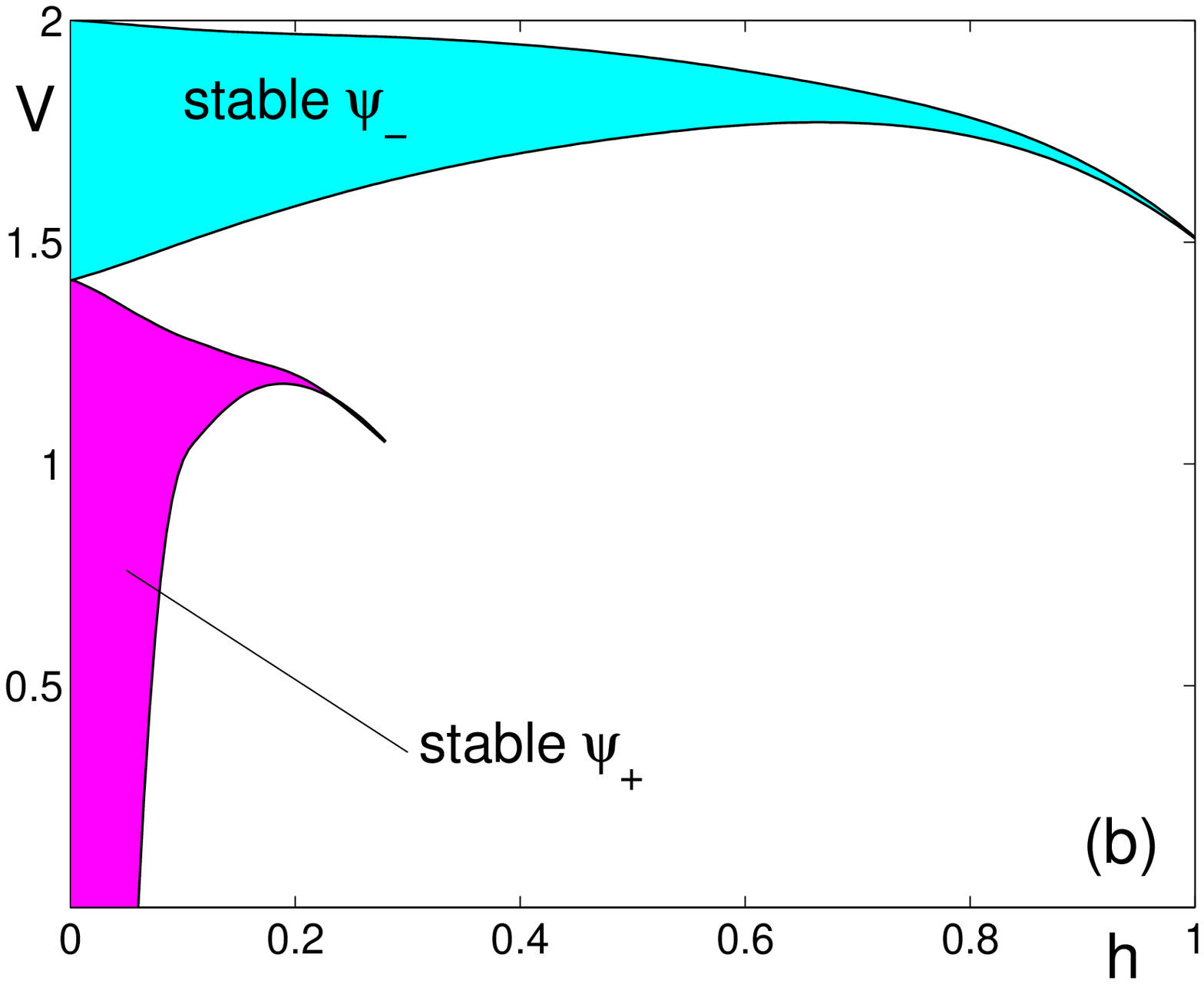}
\caption{\label{chart}
(a) The chart of the stable one-soliton solutions of the externally driven
nonlinear Schr\"odinger equation \eqref{undamped}.
Here $h$ varies from 0 to $\sqrt{2/27} \approx 0.2722$. 
(b) The corresponding attractor chart  for the parametrically driven soliton
 (adapted from \cite{Baer}).
} 
\end{figure}
\end{widetext}

In this paper, we studied stationary and moving solitons of 
the externally driven nonlinear Schr\"odinger equation,
\begin{equation}
i \psi_{t} + \psi_{xx} + 2 |\psi|^2 \psi - \psi= -h.
\label{undamped}
\end{equation}
Our continuation results  are summarised in  Fig.\ref{chart}(a)
which shows ranges of stable velocities for each value of the driving strength $h$. 

The notation $\psi_+$ and $\psi_-$ in this figure is used for the travelling 
waves obtained by the continuation of the stationary $\psi_+$ and $\psi_-$ 
solitons, respectively. The travelling soliton preserves some similarity with 
its stationary ancestor; this justifies the use of the same notation.

The uppermost curve in this figure is given by $V=c(h)$ where $c$ is the 
maximum velocity of the soliton propagation, Eq.\eqref{c}. 
This curve serves as the upper bound of the travelling $\psi_+$ soliton  existence domain.
The dotted curve demarcates the existence domain of the travelling 
$\psi_-$ soliton. For $h$ between 0 and 0.06 it coincides with the $V=c$;
for $0.06 \leq h \leq 0.2722$ it is given by $V=V_{\rm max}(h)$ where $V_{\rm max}$
is the position of the turning point in Fig.\ref{PV}(a). 

The area shaded in blue (light grey) gives the stability region of the soliton $\psi_+$
and the area shaded by purple (dark grey) is the $\psi_-$ stability domain. 
Note that the blue and purple regions partially overlap: for small $h$,
there is a range of ``stable" velocities accessible to solitons of both families. 
The light (yellow) strip inside the purple (dark grey) region
represents the stability domain of the bound state of two $\psi_-$ solitons. 

As we cross the right-hand ``vertical" boundary of the purple (dark grey) region, 
the $\psi_-$ soliton loses its stabiilty to an oscillatory mode. If we had damping in the
system, the onset of instability would correspond to the Hopf bifurcation
giving rise to a time-periodic solution. 
 In the absence of damping, 
the oscillatory instability produces an oscillatory structure with long but finite lifetime \cite{ABP}. 
These solitons with oscillating amplitude and width, travelling with oscillatory velocities,
were observed in \cite{MQB}. These
are expected to exist to the right of the purple (dark grey) region.

Where possible, we tried to emphasise the similarity of the arising
bifurcation diagrams with the corresponding diagrams  for the 
{\it parametrically\/}
driven nonlinear Schr\"odinger equation:
\begin{equation}
i \psi_{t} + \psi_{xx} + 2 |\psi|^2 \psi - \psi= h \psi^*.
\label{paramaribo}
\end{equation}
Fig.\ref{chart}(b) reproduces the soliton attractor chart for Eq.\eqref{paramaribo} \cite{Baer}. 
The structure of the stability regions in the two figures is remarkably similar. The slowly moving
solitons in the purple- (dark grey-) tinted region inherit their stability from the stationary solitons 
of the family which is stable for small $h$ (the $\psi_-$ family in the externally-driven
and the $\psi_+$ family in the parametrically-driven case). On the other hand, the 
solitons in the blue- (light grey-) shaded area are transonic (i.e. move close to $c$, velocity
of the sound waves). Their stability is due to the proximity of the nonlinear Schr\"odinger
equation to the KdV in the transonic limit \cite{BM}.

\end{document}